\documentclass[aps,prb,twocolumn,floats,showpacs,superscriptaddress]{revtex4-1}
\usepackage{graphicx,epsfig}
\usepackage{times}
\usepackage{graphics,dcolumn,bm,float}
\usepackage{bbold}
\usepackage{subfigure}
\usepackage{amssymb,amsmath,rotate,color}
\usepackage[pagebackref=false,colorlinks,linkcolor=blue,citecolor=blue,urlcolor=blue]{hyperref}
\usepackage[normalem]{ulem}

\begin{document}

\unitlength 1.5 cm
\newcommand{\be}{\begin{equation}}
\newcommand{\ee}{\end{equation}}
\newcommand{\bea}{\begin{eqnarray}}
\newcommand{\eea}{\end{eqnarray}}
\newcommand{\nn}{\nonumber}
\newcommand{\hlt}[1]{\textcolor{red}{#1}}

\title{Strong-coupling perturbative study of the disordered Hubbard model on honeycomb lattice}
\author{Alireza Habibi} 
\affiliation{Department of Physics, Sharif University of Technology, Tehran 11155-9161, Iran}
\affiliation{School of Physics, Institute for Research in Fundamental Sciences (IPM), Tehran 19395-5531, Iran}
\author{Elaheh Adibi} 
\affiliation{Department of Physics, Sharif University of Technology, Tehran 11155-9161, Iran}
\author{S. A. Jafari}
\email{jafari@physics.sharif.edu}
\affiliation{Department of Physics, Sharif University of Technology, Tehran 11155-9161, Iran}
\affiliation{Center of excellence for Complex Systems and Condensed Matter (CSCM), Sharif University of Technology, Tehran 14588-89694, Iran}

\begin{abstract} 	
We study the Anderson disordered Hubbard model on the honeycomb lattice. The Hubbard term is handled with strong-coupling perturbation theory
which encodes the Mott transition physics into a rich dynamical structure of a local self-energy. The local nature of self-energy allows us to combine
it with kernel polynomial method and transfer matrix methods. The locality of self-energy combined with the analytic nature of the 
strong-coupling perturbation theory enables us to study lattices with millions of sites. 
The transfer matrix method in the ribbon geometry is essentially free from finite size errors and allows us to perform a careful finite size scaling
of the width of the ribbon. This finite size scaling enables us to rule out the possibility of metallic phase in between the Mott and Anderson insulating phases.
We therefore find a direct transition between Anderson and Mott insulators when the disorder strength $W$ is comparable to the Hubbard interaction $U$. 
For a fixed disorder $W$, we obtain an interaction dependent nonmonotonic behavior of the localization length which reflects interaction induced 
enhancement of the localization length for weak and intermediate interaction strengths. Eventually at strong interactions $U$, the Mott localization
takes over and the localization length becomes comparable to the lattice scale. This is reminiscent of the holographic determination of the Mott state 
where the system at IR recognizes its UV lattice scale. 
\end{abstract} 
	
\maketitle

\section{Introduction\label{introduction}}
The physical properties of solids are strongly influenced by the interaction between electrons and the presence of disorder.
Localization is the most important theme in both purely disordered systems and purely correlated systems, which of course happens by 
two completely different mechanisms. In correlated systems, strong Coulomb interaction strength at half-filling leads to the gapped charge excitations due to high cost of double occupancy which is known as Mott insulator~\cite{Mott}.
On the other hand, in the presence of disorder, the eigenstates of the non-interacting system can be localized and  decay exponentially with distance due to coherent backscattering which defines the Anderson insulating state~\cite{Anderson}.

The semimetal to Mott insulator transition driven from electron-electron interaction alone on honeycomb lattice is extensively studied by various method such as quantum Monte Carlo (QMC) simulations~\cite{Sorella,Otsuka}, renormalization group methods~\cite{Herbut,Honerkamp}, cluster dynamical mean field theory (cDMFT)~\cite{He&Lu,Wu,Liebsch}, strong-coupling perturbation approach~\cite{Adibi} and so on. For disordered and non-interacting electrons on the honeycomb lattice, recent studies showed that strong long-range disorder~\cite{Zhang,Mucciolo} and short-range disorder~\cite{Aleiner,Altland,Xiong,Schubert} cause intervalley scattering which leads to Anderson localization. Furthermore, honeycomb lattice as a two dimensional lattice could be a good candidate to consider the scaling theory of localization~\cite{Abrahams}. This theory predicts that all states of the one and two dimenional system are localized at zero temperature for any finite disorder strength in the absence of electron-electron interaction and magnetic field. Schreiber and Ottomeier~\cite{Schreiber} and Fan \textit{et al.}~\cite{Fan} by using the transfer matrix method and the real-space Kubo-Greenwood method, respectively and Lee \textit{et al.}~\cite{Lee} by means of self-consistent Born approximation showed that in the presence of short-range disorder in graphene, all states are localized and obey the scaling theory of localization.  On the other hand, the results of Refs.~\onlinecite{Amini,Song} find a metal-insulator transition for uncorrelated and short-range disorder in graphene.

While the individual effects of interaction and disorder are widely examined on honeycomb lattice, the interplay of interaction and disorder on honeycomb lattice is an ambiguous and non-trivial problem. On the other hand, in real materials, both interaction and disorder are present.
So, in this paper, we set out to investigate the combined effects of the interaction and disorder on the metal-insulator transition by focusing on honeycomb lattice. 

Despite the extensive research throughout the decades on the competition of interaction and disorder on different lattices no conclusive theory has been established yet. The challenging problem of the possible existence of a metallic phase in two dimension, induced by interactions have been discussed by many authors. The metallic ground state extracted in finite size systems in two dimension is reported at Refs~\onlinecite{denteneer99,Denteneer,Chakraborty, Heidarian}. 
It was suggested that the numerically obtained metallic phase in two dimension is probably an artifact of finite sizes~\cite{Henseler}. 
The typical numerical methods such as QMC~\cite{Meng,MoreoQMC}, exact diagonalization~\cite{noceED}, cDMFT~\cite{He&Lu,ParkCDMFT}, variational cluster approximation~\cite{SekiVCA,Balzer} etc that are routinely
used to handle the interaction part suffer from severe size limitations rooted in exponential growth of the Hilbert space.
It would be therefore desirable to employ an analytic procedure to handle the interaction part. 
To better understand the puzzles on the interplay of interactions and disorder, in this paper we use a method which does not suffer from 
such severe finite size effects, which will in turn enable us to perform a reliable finit size scaling. 

Let us briefly introduce the method we employ to perturbatively solve the interaction part.
We employ the so called strong-coupling perturbation theory~\cite{Senechal98,Senechal2000} which can be used to calculate the Green's function 
of the Hubbard Hamiltonian analytically for infinite lattice. In this method, the inter-site hopping $t$ is considered as the perturbation parameter, 
so that one can carry out the perturbation expansion about the atomic limit in powers of $t/U$ where $U$ is the Hubbard interaction strength. 
Since the typical values of critical $U/t$ needed for Mott transition are $\sim 3$, even a low-order perturbation treatment in $t/U\sim 1/3$ can
capture the Mott aspect spectacularly. The highly non-trivial information on Mott physics is encoded in the dynamical self-energy that can be 
analytically computed in this method.
This self-energy is local and therefore it can be naturally incorporated to disordered situations. This procedure is 
free from any finite size artifacts on the Hubbard side.  
Placing non-trivial (and local) self-energies on a lattice allows to combine it with on-site Anderson disorder (measured by the width $W$ of the on-site energy) 
which then can be handled numerically in a very efficient way. Employing the kernel polynomial method (KPM) allows us to calculate the density of state (DOS) for 
disordered interacting system 
with millions of lattice sites in the real space. In this method, any spectral function is expanded in terms of Chebyshev (or any other complete set of orthonormal) 
polynomials, where the expansion coefficients are obtained through an efficient recursion relation involving matrix elements of the Hamiltonian in stochastically 
sampled states~\cite{kpm,Habibi}. The central result obtained from DOS is that in presence of disorder there is a direct transition from
Anderson insulator to Mott insulator which takes place at a critical interaction $U_c\approx W$.
To get further insight into the behavior of the disordered Hubbard model, we utilize the transfer matrix method~\cite{MacKinnon1,MacKinnon2} 
to compute the localization length. The finite size scaling analysis of the localization length can conclusively determine whether system is metal or Anderson insulator. 
The localization length is considered as the relevant scale which determines the transport properties of the system. 
In agreement with previous numerical results~\cite{Henseler,Atkinson,Srinivasan} reported for Anderson-Hubbard model, for a fixed
large disorder strength $W$, by increasing $U$ the localization length increase and after reaching a maximum starts to decrease. 
The increase in the localization length can be attributed to the screening of disorder by interactions.
Our finite size scaling shows that even the maximal localization length indeed correspond to Anderson
insulating state. 
This enables us to rule out a putative metallic state in between the Anderson and Mott insulating state. 

The rest of this paper is organized as follows. We begin by introducing the Anderson-Hubbard model to study the interacting disorder system and then briefly reviewing the strong-coupling approach in Sec.~\ref{Model and Method}. Next in Sec.~\ref{Results}, we present our results for interplay of interaction and disorder. Finally, in Sec.~\ref{Concluding Remarks}, we end up with some concluding remarks. The article is accompanied by three appendices which first present the one-point correlation function of the atomic-limit of the Hamiltonian and afterwards in two other appendices we describe the KPM and transfer matrix method.

\section{Model and Method\label{Model and Method}}
We study the disordered interacting system by the Anderson-Hubbard model which is given by the following Hamiltonian,
\bea
&&H=H_0+H_1,\label{Hamiltonian}\\
&&H_0=U \sum_{i} n_{i\uparrow} n_{i\downarrow}-\mu\ \sum_{i,\sigma}\ n_{i\sigma}+\sum_{i,\sigma}\ \epsilon_i\ n_{i\sigma},\label{unperturbed-H}\nn\\
&&H_1=\sum_{ij,\sigma}\ V_{ij}\ (c_{i\sigma}^{\dagger}\ c_{j\sigma}+H.c.)\label{perturbation},\nn
\eea
where $H_0$ accounts for interaction and disorder energy, and $H_1$ for kinetic energy. Also $c_{i\sigma}^{\dagger}$ and $c_{i\sigma}$ are, respectively, the fermionic creation and annihilation operators of the particle with spin $\sigma=\uparrow, \downarrow$ on the lattice site $i$, $n_{i\sigma}=c_{i\sigma}^{\dagger}\ c_{i\sigma}$ measures the occupation of site $i$ with an electron of spin $\sigma$, $V_{ij}$ is the hopping matrix element between sites $i$ and $j$, $U$ is the on-site Hubbard repulsion and $\mu$ is the chemical potential. The disorder affects system by local term in $H_0$ which is parameterized with a random potential $\epsilon_i$ with a box probability distribution $P(\epsilon_i)=\Theta(W/2-|\epsilon_i|)/W$, where $\Theta$ is the step function. The parameter $W$ is a measure of the disorder strength.

In what follows, we briefly describe the strong-coupling perturbation theory~\cite{Senechal98}. Considering $H_0$ and $H_1$ in Hamiltonian~(\ref{Hamiltonian}) 
as the unperturbed and perturbed parts respectively, the partition function at temperature $T=1/\beta$ in the path-integral formalism is written as,
\bea
Z&=&\int [d\gamma^{\star} d\gamma]\ \exp\bigg[ -\int_{0}^{\beta}d\tau\bigg\lbrace \sum_{i\sigma} \gamma^{\star}_{i\sigma}(\tau)\ \partial_\tau\ \gamma_{i\sigma}(\tau)\nn\\
&+&H_0(\gamma^{\star}_{i\sigma}(\tau),\gamma_{i\sigma}(\tau))+\sum_{ij\sigma}\gamma^{\star}_{i\sigma}(\tau)\ V_{ij}\ \gamma_{j\sigma}(\tau) \bigg\rbrace\bigg], 
\eea
where $\gamma$ and $\gamma^{\star}$ denote the Grassmann fields in the imaginary time $\tau$. 

In the absence of the Wick's theorem for the unperturbed Hamiltonian, employing the standard perturbation theory is not straightforward. 
The Wick's theorem is borough to life by applying the following Hubbard-Stratonovich transformation,
\bea
&&\int [d\psi^\star d\psi] \exp\bigg[\int_{0}^{\beta} d\tau \sum_{i\sigma} \bigg\lbrace\sum_j \psi^\star_{i\sigma}(\tau) (V^{-1})_{ij} \psi_{j\sigma}(\tau)\nn\\
&&\qquad\qquad+\psi^\star_{i\sigma}(\tau)\gamma_{i\sigma}(\tau)+\gamma^\star_{i\sigma}(\tau)\psi_{i\sigma}(\tau)\bigg\rbrace\bigg]\nn\\
&&=\det(V^{-1})\exp\bigg[-\int_{0}^{\beta} d\tau \sum_{ij\sigma} \gamma^\star_{i\sigma}(\tau)\ V_{ij}\ \gamma_{j\sigma}(\tau) \bigg],
\eea
where $\psi_{i\sigma}(\tau)$ and $\psi^\star_{i\sigma}(\tau)$ are the auxiliary Grassmann fields. Actually, by means of this transformation, we can rewrite the partition function up to a normalization factor as,
\bea
Z=\int[d\psi^\star d\psi] \exp\bigg[-\bigg\lbrace S_0[\psi^\star,\psi]+\sum_{R=1}^{\infty} S^R_{int}[\psi^\star,\psi]\bigg\rbrace \bigg].\label{rewritten-Z}
\eea
As can be seen, the new representation of the partition function is in terms of the auxiliary fermions. $S_0[\psi^\star,\psi]$ is the free auxiliary fermion action given by the inverse of the hopping matrix of 
original fermions, 
\bea
S_0[\psi^\star,\psi]=-\int_{0}^{\beta}\ d\tau \sum_{ij\sigma} \psi^\star_{i\sigma}(\tau)\ (V^{-1})_{ij}\ \psi_{j\sigma}(\tau),
\eea
and $S^{R}_{int}[\psi^\star,\psi]$  is an infinite number of interaction terms given by,
\bea
 &&S^{R}_{int}[\psi^\star,\psi]=\frac{-1}{(R!)^2}\sum_{i}\sum_{\lbrace\sigma_l\sigma^{\prime}_l\rbrace} \int_{0}^{\beta}\ \prod_{l=1}^{R}\ d\tau_l d\tau'_l\nn\\ &\times &\psi_{i\sigma_1}^{\star}(\tau_1)\ldots\psi_{i\sigma_R}^{\star}(\tau_R)\psi_{i\sigma'_R}(\tau'_R)\ldots\psi_{i\sigma'_1}(\tau'_1)\nn\\
&\times &\bigg\langle \gamma_{i\sigma_1}(\tau_1)\ldots\gamma_{i\sigma_R}(\tau_R)\gamma^{\star}_{i\sigma'_R}(\tau'_R)\ldots\gamma^{\star}_{i\sigma'_1}(\tau'_1)\bigg\rangle_{0,c}.\label{interaction-Z}
\eea
$\langle \gamma_{i\sigma_1}(\tau_1)\ldots\gamma_{i\sigma_R}(\tau_R)\gamma^{\star}_{i\sigma'_R}(\tau'_R)\ldots\gamma^{\star}_{i\sigma'_1}(\tau'_1)\rangle_{0,c}$ represents the connected correlation function. In the diagrammatic representation, this correlation function denote a $2R$ apices vertex which is attached to $R$ 
incoming ($\psi$) and $R$ outgoing ($\psi^\star$)  auxiliary fermions. 

In the partition function of the auxiliary fermions~(\ref{rewritten-Z}), the free propagator is given by matrix $V$. So, we can apply the  Wick's theorem to consider the interaction term~(\ref{interaction-Z}) perturbatively and  calculate the self-energy of the auxiliary fermion ($\Gamma$). Finally, the Green's function of the original fermions is expressed by,
\bea
G=(\Gamma^{-1}-V)^{-1}\label{green'sfunction}.
\eea
For more details on the strong-coupling approach, see Ref. \onlinecite{Senechal2000}.

\section{Results\label{Results}}

We consider the honeycomb lattice in which $V_{ij}=-t$ if $i,j$ are nearest neighbour sites and is zero otherwise. 
Also, throughout the paper we choose $t=1$ as the energy unit.
In realistic graphene the energy scale is set by $t\sim 2.8$ eV~\cite{Reich}. 
We are interested in half-filling which is defined by $\overline{\langle n_{i\sigma}\rangle}=1/2$ where $\langle n_{i\sigma}\rangle$ denotes the 
mean occupation of each site for a given spin projection and a fixed realization of randomness which is given by following equation,
\bea
\langle n_{i\sigma}\rangle=\frac{e^{\beta(U/2-\epsilon_i)}+e^{-2\beta\epsilon_i}}{1+2e^{\beta(U/2-\epsilon_i)}+e^{-2\beta\epsilon_i}},
\label{nisigma.eqn}
\eea
where $\epsilon_i$ are random on-site energies distributed in a box of width $W$. The bar in $\overline{\langle ...\rangle}$ denotes averaging over 
realizations of disorder. 
In the absence of $\epsilon_i$ term in the Hamiltonian~(\ref{Hamiltonian}), the half-filling is simply realized by setting the chemical potential $\mu=U/2$. 
In presence of the disorder term the plot of $\langle n_{i\sigma}\rangle$ as a function of the chemical potential $\mu$ 
at zero temperature consists in three plateaus corresponding to values of $0,0.5$ and $1$. 
For $W<U$, only the portion of plateau corresponding to $0.5$ is realized and therefore $\mu=U/2$ establishes the half-filling. 
For $W>U$, the occupation $\langle n_{i\sigma}\rangle$ in addition to $0.5$ has a chance to pick up $0,1$ as well. 
However, due to the symmetry of Eq.~\eqref{nisigma.eqn} around $\mu=U/2$, the chance of realizing occupation of $0$ and $1$ is equal. 
Therefore, again the previously mentioned chemical potential specifies the half-filling. So, in the presence of any disorder, we still use the $\mu=U/2$.

We treat the Mott-Hubbard aspects within the leading order of the strong-coupling perturbation theory
which is already capable of capturing the Mott physics. At this order, the dynamical self energy 
of auxiliary fermions is expressed by the one-point connected correlation function (for derivation see Appendix~\ref{cumulant}).
So, in this limit the self-energy of the auxiliary fermions at half-filling for each spin is given by
\bea
\Gamma_{ij}(i\omega)=\Big(\frac{1-\langle n_{i}\rangle}{i\omega-\epsilon_i+U/2}+\frac{\langle n_{i}\rangle}{i\omega-\epsilon_i-U/2}\Big)\ \delta_{ij}\label{self-energy},
\eea
where $i\omega$  denotes to Matsubara frequency and $\delta_{ij}$ is Kronecker delta. Note that according to Hamiltonian (\ref{Hamiltonian}), 
in the absence of symmetry breaking, there is no difference between $\langle n_{i\uparrow}\rangle$ and $\langle n_{i\downarrow}\rangle$, so we just use $\langle n_{i}\rangle$ for mean occupation.

\subsection{Mott gap equation in disordered systems}
\begin{figure}[t]
    \centering
    \includegraphics[width=1\linewidth]{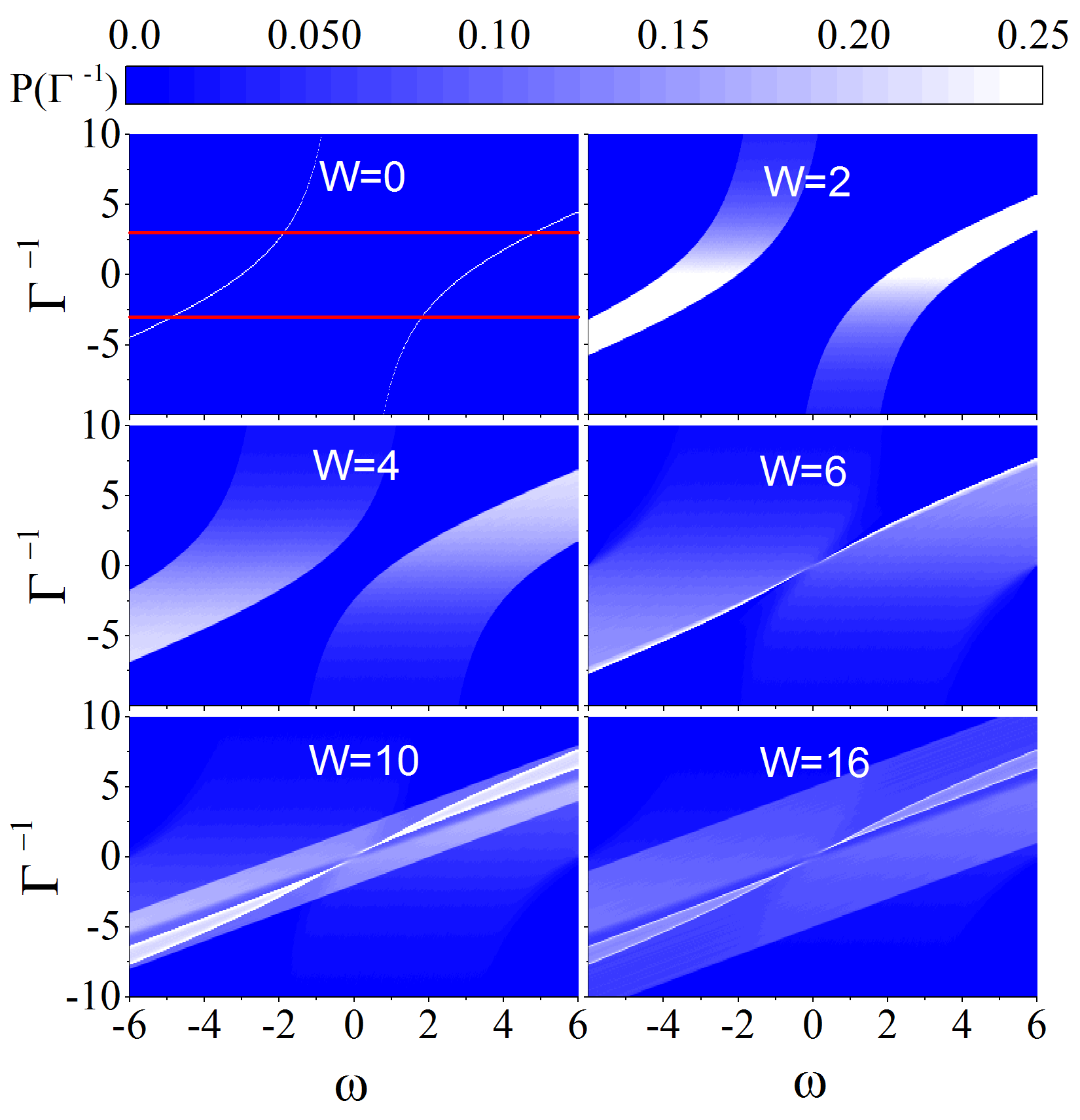}
    \caption{(Color online) The distribution of $\Gamma^{-1}(\omega)$ for different disorder strength $W$ at $U=6$. 
    Red lines in the upper left panel show the critical values $\Gamma^{-1}=\pm 3$ beyond which the Mott state is realized. } 
    \label{omega-gamma-dis}
\end{figure}
As presented in detail in Ref.~\onlinecite{Adibi}, the strong-coupling perturbation theory enables us to 
set up a {\em gap equation} for the Mott state. The DOS of the clean interacting electrons on the honeycomb lattice was found to be 
$\rho(\omega)=\rho_0(\Gamma^{-1}(\omega))$ where $\rho_0$ denotes the DOS of non-interacting electrons. 
Due to threefold coordination of the honeycomb lattice, $\rho_0$ is nonzero if and only if the absolute value of its argument 
does not exceed $3$. Therefore, the criterion $\left|\Gamma^{-1}(\omega+i0^+)\right|_{\omega=0}\ge 3$ determines the Mott state
in clean system. In disordered systems, level repulsion increases the half-bandwidth $3$. The appropriate generalization
of this criterion for disordered systems will be 
\be
   \left|\overline{\Gamma^{-1}(\omega+i0^+)}\right|_{\omega=0}\ge B^0_W,
   \label{mottgapeq}
\ee
where $B^0_W$ is the half-bandwidth of the {\em non-interacting but disordered} system  which will now depend on the disorder strength $W$. 
This relation simply expresses the disorder averaged version of the condition that the denominator of Eq.~\eqref{green'sfunction} does not pick a pole at $\omega=0$.
For the clean system one obviously gets the half-bandwidth of clean non-interacting system $B^0_{W=0}=3$~\cite{Adibi}.

Now let us see how does $\Gamma^{-1}$ -- which is related to the self-energy of physical electrons -- respond to Anderson disorder. 
As can be seen from Eq.~\eqref{self-energy}, the self-energy $\Gamma$ of auxiliary fermions parametrically depends on the random 
on-site energies $\epsilon_i$. Therefore the random distribution of $\epsilon_i$, induces a distribution of $\Gamma$ which will now be a dynamical 
distribution as it depends on frequency $\omega$. This has been plotted in Fig.~\ref{omega-gamma-dis} 
which shows distribution of $\Gamma^{-1}(\omega)$ at all frequencies for various values of disorder strength $W$, and
a fixed $U=6$ Hubbard interaction.
Taking advantage of the criterion~\eqref{mottgapeq}, the important feature obtained from Fig.~\ref{omega-gamma-dis} 
is that at frequencies where the distribution of $\Gamma^{-1}$ takes an average value between $-B^0_W$ and $B^0_W$, the 
{\em interacting} DOS at that frequency is non-zero.
Owing to the particle-hole symmetry, the possible Mott-Hubbard gap opens up at $\omega=0$. So we focus on zero frequency.
As can be seen in the absence of disorder, $W=0$ (top left panel in Fig.~\ref{omega-gamma-dis}), $\Gamma^{-1}$ 
is distributed on a line of zero width. Also, this line distribution at $\omega=0$ already falls outside the range of $(-B^0_0,+B^0_0)$. 
Therefore, the interacting DOS is gapped for $W=0$ and $U=6$, and therefore the system is in its Mott insulating phase. 
By turning the disorder on, the distribution of $\Gamma^{-1}$ start to broaden and as demonstrated for $W \geq 6$, the distribution of  
$\Gamma^{-1}$ will move most of its weight to $\omega=0$, such that its average at $\omega=0$ falls in the non-interacting bandwidth specified by $B^0_W$. 
This means that large enough disorder strength closes the Mott gap and the system becomes Anderson insulator. 
By further increasing the disorder strength, -- specially the $\omega\approx 0$ portion of -- $\Gamma^{-1}(\omega)$ distribution
becomes more concentrated in the non-interacting bandwidth $B^0_W$.

\subsection{Competition between Anderson localization and Mottness}
\begin{figure}[t]
    \centering
    \includegraphics[width=0.8\linewidth]{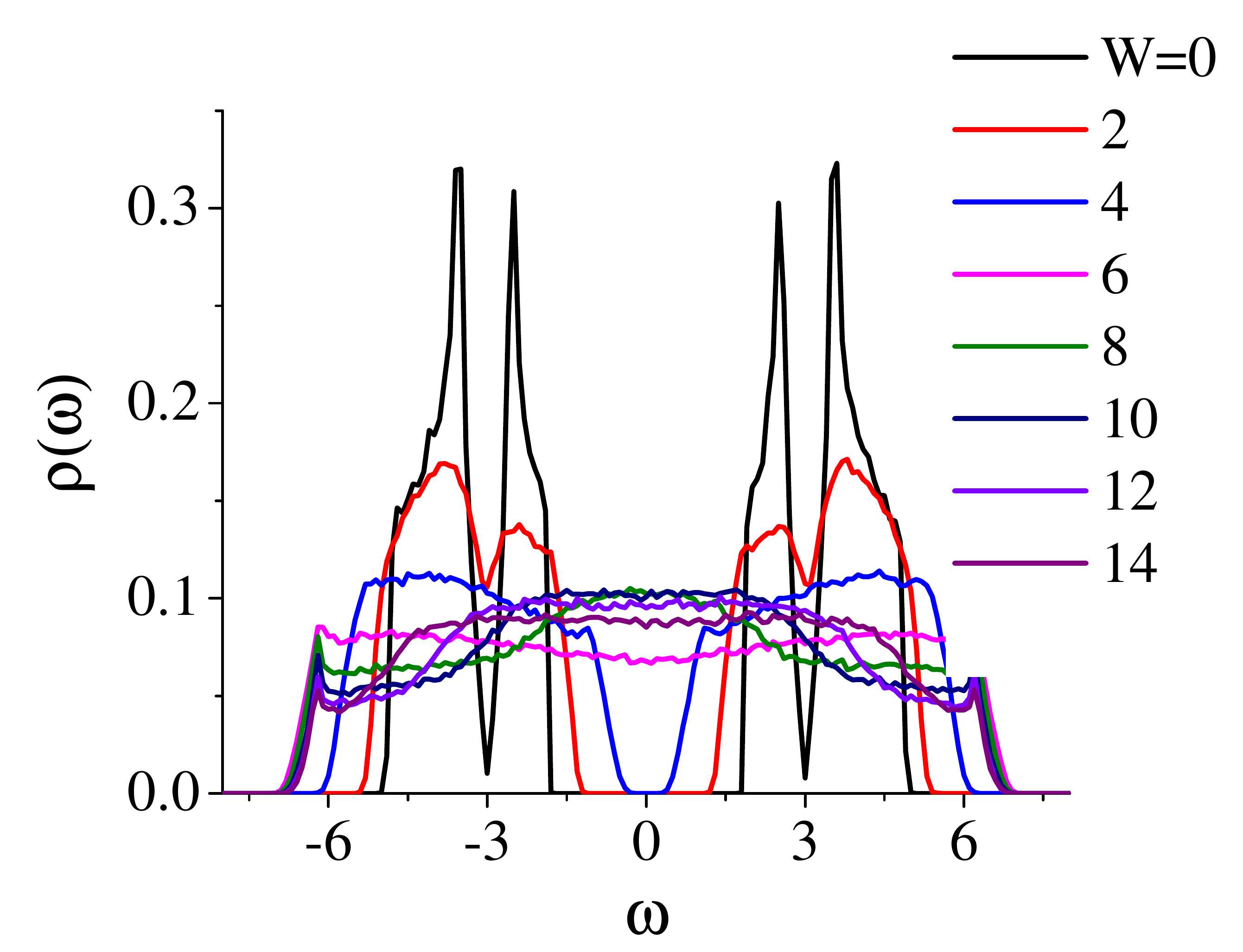}
    \caption{(Color online) The evolution of the disorder-averaged DOS as a function of $W$ for fixed $U=6$ at half-filling and zero temperature.}
    \label{DOS-U=6}
\end{figure}

To gain a better understanding of what explained for Fig.~\ref{omega-gamma-dis}, instead of expressing the 
condition for picking up a non-zero density of states at $\omega=0$, let us actually calculate the relevant
trace (Tr) in Eq.~\eqref{green'sfunction}. This can be efficiently done with the KPM~\cite{kpm}. But in the present 
case due to nonlinear dependence of $\Gamma^{-1}(\omega)$ on $\omega$, it requires a trick which has been explained in 
the appendix~\ref{KPM}. Doing the summation required in the Tr of Eq.~\eqref{green'sfunction} in Fig.~\ref{DOS-U=6} we
obtain the disorder-averaged DOS at half-filling and zero temperature in different disorder strengths and $U=6$ for 
a lattice with $500 \times 500$ sites. 
Note that the present method being strong-coupling expansion in $t/U$ works better for lager $U$. 
We have benchmarked the $W=0$ (black curve) DOS of our KPM algorithm against Ref.~\onlinecite{Adibi}. 
As can be seen in Fig.~\ref{DOS-U=6} for $U=6$ in the absence of disorder, the system has already a Mott gap as expected from Refs.~\onlinecite{Sorella,Adibi}. 
As we pointed out, by turning on the disorder, it gradually broadens the DOS which eventually closes the gap at the disorder strength of $W \approx U$. 
The evolution of a clear Mott gap to a pseudogap and subsequently filling the gap, destroys the Mott phase.
Therefore we will be dealing with situations where there are states present at the Fermi level.
Now the question is whether these states are Anderson localized or extended?

The remarkable feature of DOS is that since for non-zero disorder the Mott gap is already suppressed, one requires 
much larger $U$ to restore the Mott gap of the clean ($W=0$) limit. This means that
the disorder affect the Mott transition by pushing it to larger interaction strength as also reported in Refs.~\onlinecite{Byczuk,Haase}. 

\begin{figure}[t]
    \centering
    \includegraphics[width=1\linewidth]{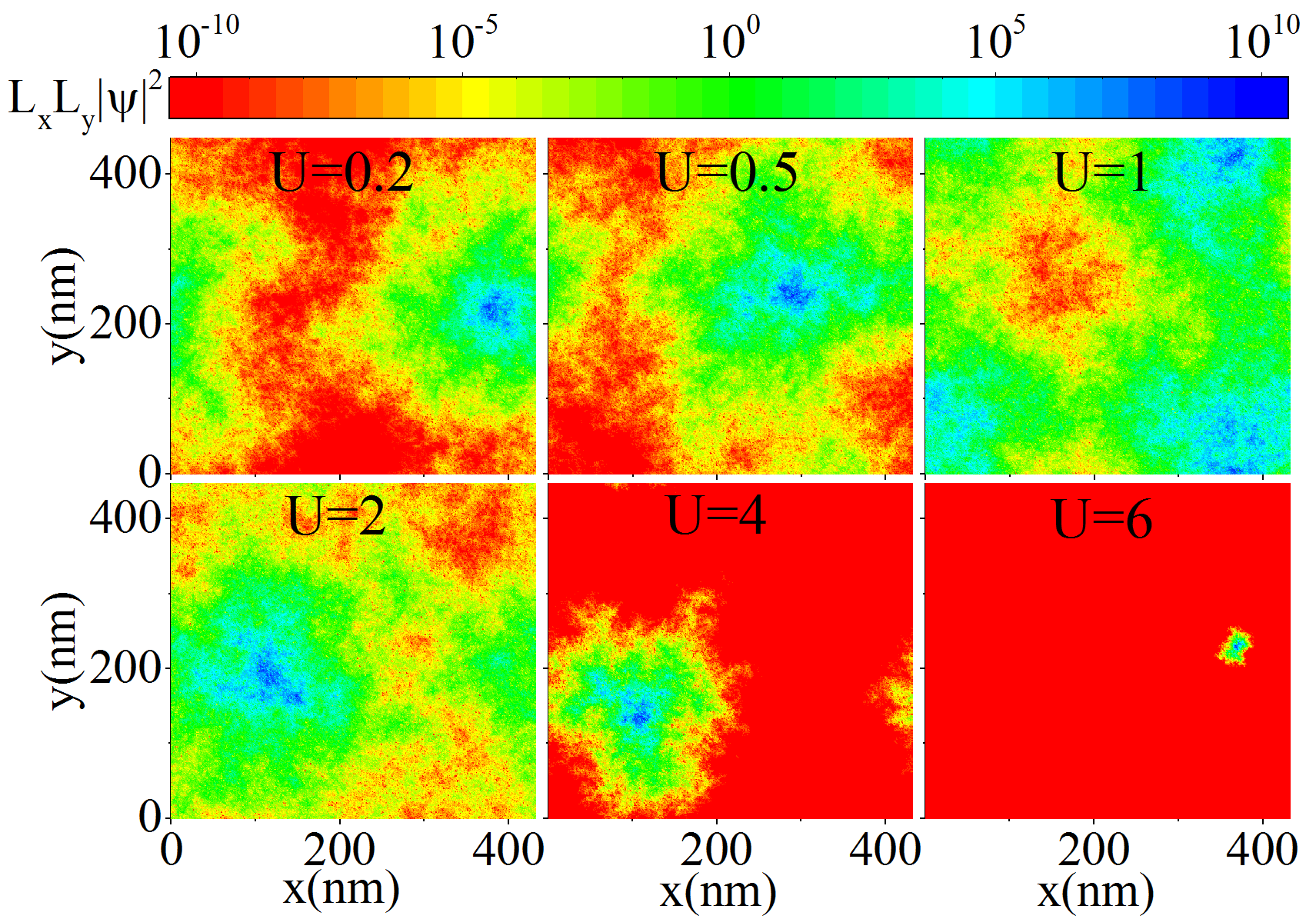}
    \caption{(Color online) Interaction dependence of the charge density (arbitrary units) at the Fermi energy for $W=6$
    for the honeycomb lattice of graphene. The size of the system is indicated in nano-meter. The eigenstates are localized for 
    small interaction strengths $U=0.2,0.5$ and very large $U=4,6$. In the intermediate strengths, despite apparent percolation 
    of the charge density which suggests a metallic state, it is not enough to specify the nature of the intermediate phase.}
    \label{U-psi2}
\end{figure}

To characterize the nature of the expected phases of the model, let us employ the exact diagonalization to generate a snapshot of the 
charge density (wave function squared) at Fermi energy. This is shown in Fig.~\ref{U-psi2} for fixed disorder strength $W=6$. 
As  illustrated in this figure, the system is Anderson localized for $U=0.2$ and $U=0.5$, as the charge density consists in 
disconnected puddles. By increasing the interaction, at $U=1$ and $U=2$ it appears that the charge puddles percolate and one is
tempted think that these values of $U$ correspond to an intermediate conducting phase. In Refs.~\onlinecite{denteneer99,Denteneer,Chakraborty} using QMC method and \onlinecite{Heidarian} by self-consistent Hartree-Fock calculations, 
the authors identify the apparent percolating charge density with metallic phases. However, we will shortly show that this is 
an artifact of very small sizes. A careful finite size scaling based on transfer matrix method will show that the system is still in the Anderson localized phase. 
Upon further increase of the interaction in Fig.~\ref{U-psi2}, we again have Anderson localized state at $U=4$ and $U=6$. 
If we continue to increase the Coulomb interaction, the Mott gap appears, and there will be no states at the Fermi level ($\omega=0$). 

\subsection{Characterization of intermediate phase with transfer matrix}
Let us return to the metallic-looking phase for $U\sim 1-2$. As pointed out, even the sizes indicated in
Fig.~\ref{U-psi2} are not enough to judge whether the system is Anderson localized, or the wave functions 
are conducting. To make a conclusive judgment about the nature of this intermediate phase, we need
to go to much larger sizes for which exact diagonalization method is handicapped. 
To overcome this problem, we employ the transfer matrix method explained in Appendix~\ref{TTM}.

\begin{figure}[t]
    \centering
    \includegraphics[width=0.99\linewidth]{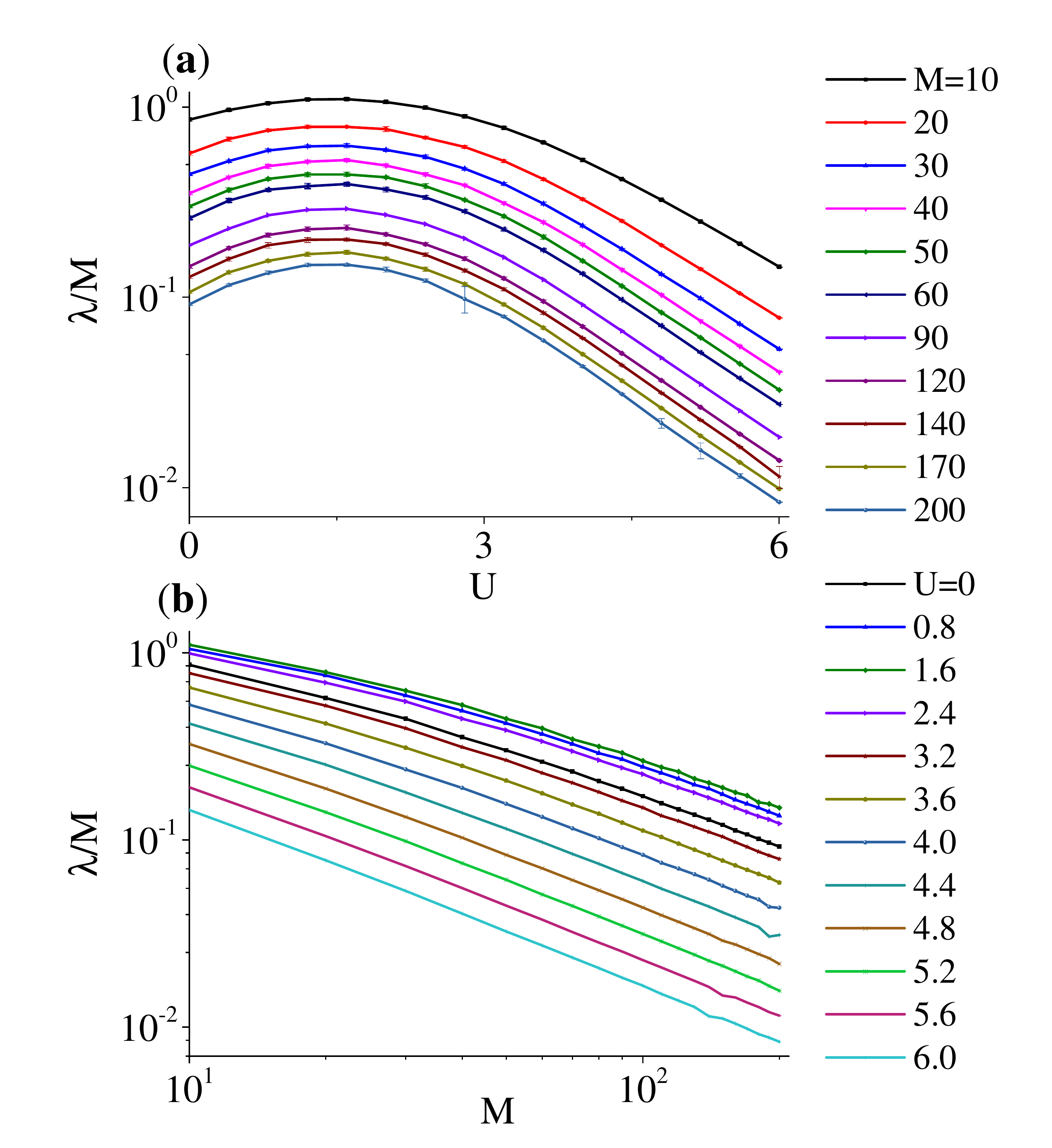}
    \caption{(Color online) The localization length normalized to the width in ribbon geometry, at the Fermi level as a function of (a) 
    interaction strength $U$ for various ribbon widths $M$ indicated in the legend,  and (b) the ribbon width $M$ for various Hubbard $U$ values. 
    In both cases, the disorder strength is fixed at $W=6$ and lattice length $L=10^5$. 
    The apparent percolation structure in Fig.~\ref{U-psi2} for $U\sim 1-2$ corresponds to a maximum of localization length in panel (a) above.
    However, the finite size scaling in (b) indicates that scaling behavior of the normalized localization length 
    in the intermediate values $U\sim 1-2$ is no different from the other values.
    }
    \label{lambda}
\end{figure}
In Fig.~\ref{lambda}, we plot the localization length normalized to the width in ribbon geometry, $\lambda/M$, at zero energy for lattice with length $L=10^5$ 
at disorder strength $W=6$. 
The remarkable feature in using the strong-coupling approach is that we can study the large lattices in contrast to 
numerical methods such as exact diagonalization which suffer from finite size limitations. 
In panel (a) the normalized localization length is plotted as a function of Hubbard $U$ for various values of the
ribbon width, $M$, indicated in the legend. In panel (b), we plot the normalized localization length as a function of $M$ for various values of $U$. 
As can be seen in panel (a), the normalized localization length reaches a maximum value for all the ribbon widths. 
This maximum takes place for $U\approx 1.39$. This indeed corresponds to the percolation structure in Fig.~\ref{U-psi2}.
This behavior is in agreement with previous works that compute the disorder-averaged inverse participation ratio in Refs.~\onlinecite{Atkinson,Henseler}. 
As far as Fig.~\ref{U-psi2} is concerned, it is tempting to interpret the intermediate regime $U\sim 1-2$ as metallic state.
However as can be seen in panel (b) of Fig.~\ref{lambda}, the scaling behavior of the normalized localization length with the ribbon width
$M$ is identical for all $U$ values indicated in the figure. Therefore the intermediate regime, $U\sim 1-2$ is not different from the 
other values of $U$ as long as there is no Mott gap in the spectrum. They all correspond to the Anderson localized state. 

\begin{figure}[t]
    \centering
    \includegraphics[width=0.8\linewidth]{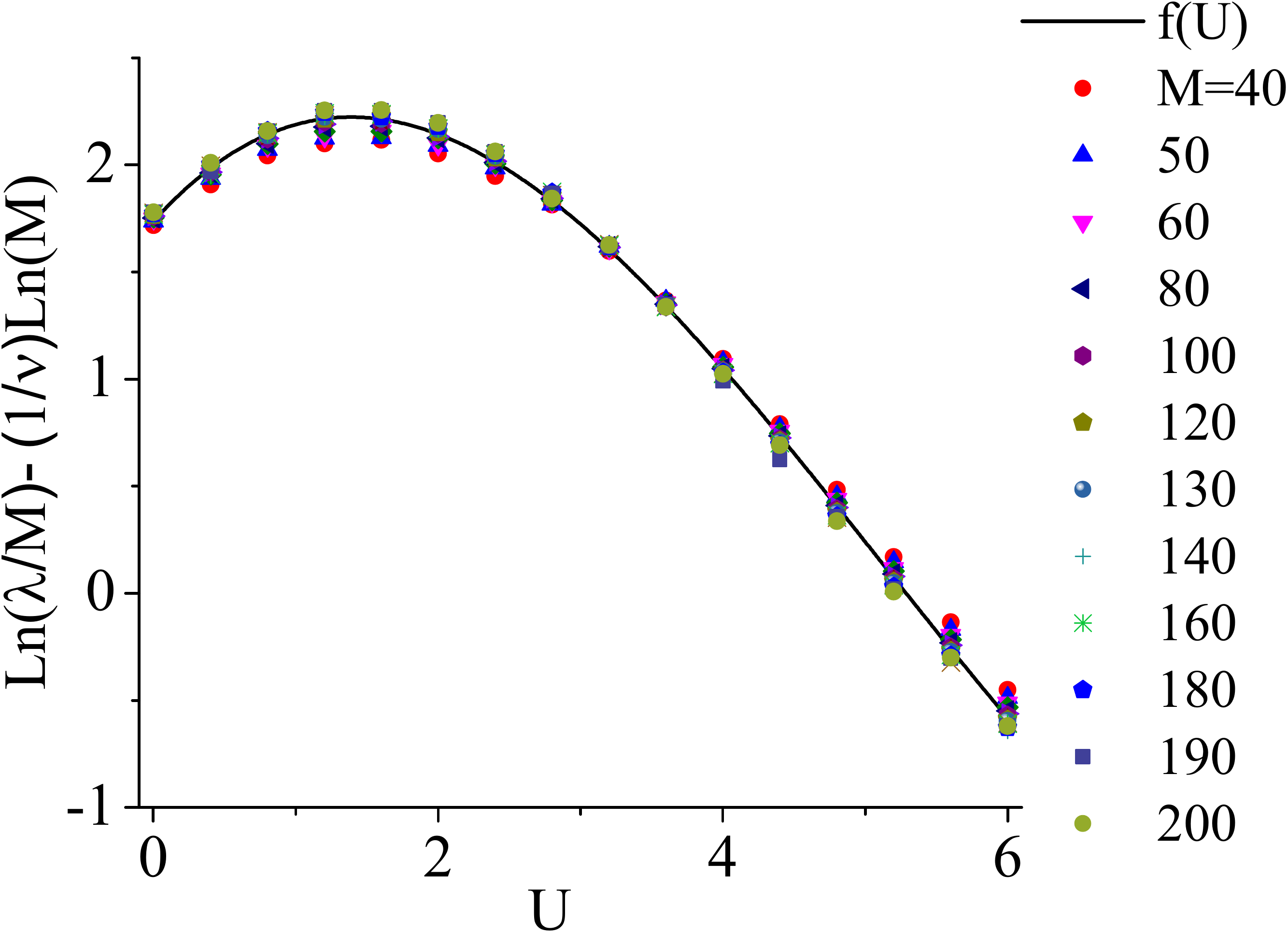}
    \caption{(Color online) Scaling analysis corresponding to Eq.~\eqref{scaling.eqn} on data of Fig.~\ref{lambda}. The
    exponent $\nu= 1.38\pm  0.16$ is has a very weak dependence on $U$. 
    This figure is produced for disorder strength, $W=6$.
    }
    \label{scaling.fig}
\end{figure}

As can be seen in panel (b), for almost all values of $U$, (up to $U\approx W$) the log-scale plots of $\lambda/M$ versus $M$ appear to be parallel 
lines. This suggests a relation of the form
\be
   \frac{\lambda}{M}=M^{-1/\nu} e^{f(U)},
   \label{scaling.eqn}
\ee
In Fig.~\ref{scaling.fig} we have performed this scaling where the solid line is the form of function $f(U)$.
There is slight blurring in the data, which can be accounted by a very weak dependence of the scaling 
exponent $\nu$ on $U$. In this language the intermediate region $U\sim 1-2$ simply corresponds to a maximum
of the coefficient $\exp(f(U))$ and does not changed the localized nature of the Fermi level wave functions
which is given by almost $U$ independent value 
$\nu= 1.38\pm  0.16$. 
The scaling function $f$ formalizes the idea of the screening of disorder by Hubbard interaction $U$. 
Starting from $U=0$, by increasing $U$, the localization length increases until it reaches a maximum
at $U_{0}\approx 1.39$. This tendency of the Fermi level states to become less localized
can be interpreted as the screening of the disorder by interactions. Beyond $U_0$, the screening is saturated,
and hence the localization length decreases again, which is reflected with in the decreasing behavior of $f(U)$. 

When does the Anderson localized phase ends? According to Fig.~\ref{scaling.fig}, the localization 
after reaching a maximum that corresponds to saturated screening, starts  to fall off. Therefore by further
increasing the Hubbard $U$, the wave functions at the Fermi level will become more and more localized. 
But the localization is driven by the Mottness tendency and is controlled by the Hubbard $U$. Ultimately,  
when the localization length $\lambda$ reaches the lattice scale, i.e. $\lambda\sim 1$ the localized
wave function will recognize its ultraviolet (UV) lattice and therefore the low-energy states at the Fermi level 
now have a sense of lattice. Indeed holographic Mott insulator precisely corresponds to appearance of
the UV lattice scale in the infrared (i.e. at Fermi level)~\cite{Hartnoll}. Beyond this point where the system becomes
Mott insulator, there will be no states at zero energy. 

\begin{figure}[t]
    \centering
    \includegraphics[width=0.8\linewidth]{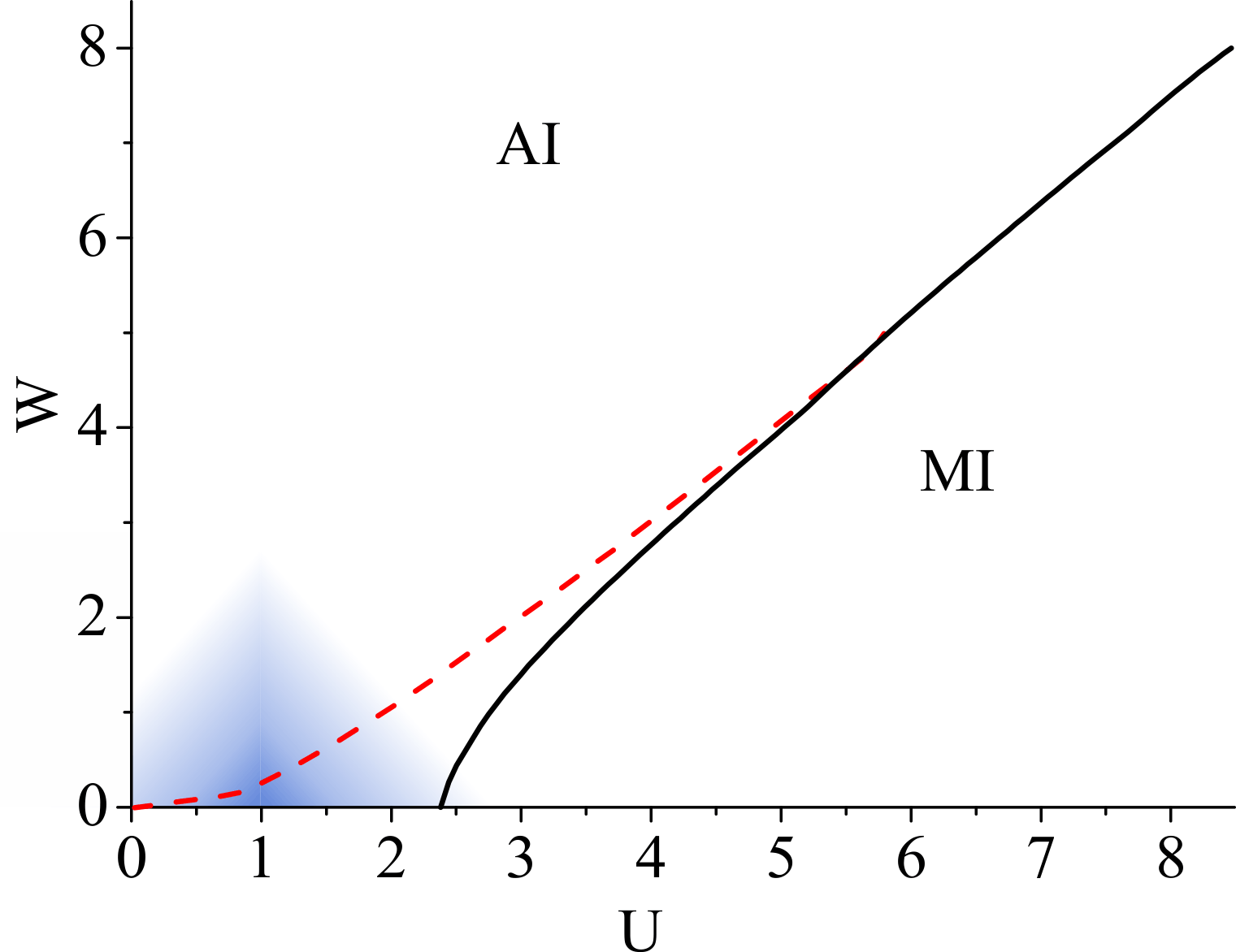}
    \caption{(Color online) Phase diagram in the $UW$ plane for the Anderson-Hubbard model on the honeycomb lattice at half-filling and zero temperature. AI and MI refer to Anderson and Mott insulator, respectively. Dashed line indicates where the spectral gap appears, while the solid line represents where extrapolated
    gap from the Mott side closes. 
    }
    \label{phasediagram}
\end{figure}

The strong-coupling method used to handle the Hubbard part of the Hamiltonian is based on large $U$ limit.
At the leading order of $t/U$ considered in this paper, the method is expected to work better at larger $U$.
Indeed at small $U$, any value of $U$ produced a nominal spectral gap. In Fig.~\ref{phasediagram} this has been
denoted by the dashed line. This is a known pathology of this method. The solution is to find out the gap for
large values of $U$, and then to extrapolate the gap trend~\cite{Adibi,Sahebsara}.  
This gives a better estimate of the critical $U$ needed
for Mott transition. This idea can also be applied to disordered problem. For a given $W$, we start from large interaction strength $U$
and extrapolate the gap to smaller values of $U$. This gives the solid line in Fig.~\ref{phasediagram}. 
For large enough $U$ where the dashed and solid boundaries in Fig.~\ref{phasediagram} agree, the transfer matrix computation of 
the localization length works very well, and the onset of Mott gap opening is where the localization length becomes of the lattice
scale. However, by reducing $W$, the transfer matrix method starts to see the lattice when it hits the dashed line.
This limit however, does not coincide with the onset of true Mott gap (solid line). The reason is that the strong coupling expansion
being an expansion in powers of $t/U$ is reliable for large enough $U$, and therefore the extrapolation of the gap from the large $U$ (Mott)
side is more reliable. Therefore for region of small $U$ and small $W$
indicated by shaded area -- around $U/t\sim 1$ where can not be reached by perturbation from neither sides -- 
the present method can not determine whether there is any conducting phase between the
Anderson and Mott insulator or not. But for large enough $U$ and $W$, the present method supported by finite size scaling, completely
rules out the possibility of a conducting phase between Anderson insulator and Mott insulator. Therefore the conclusion is that for large enough
Hubbard $U$, there is a critical $W$ beyond which the system directly transforms from Mott localized phase to Anderson localized phase. 
The equivalent picture if one walks along a fixed $W$ line is that for a fixed strong disorder $W$, beyond a critical $U$
there will be a direct transition from Anderson insulating state to Mott insulating state. 
The phase diagram of the Coulomb interaction $U$ vs the strength of disorder $W$ at zero temperature and half-filling in
Fig.~\ref{phasediagram} shows that two insulating phases are separated roughly at $U \approx W$ which is consistent with results of a 
self-consistent study~\cite{Henseler} and DMFT~\cite{Atkinson} in two dimension. In three dimensions similar picture is obtained 
by QMC method~\cite{Hatsugai}. Infinite dimensional version of DMFT gives a similar picture~\cite{Byczuk}.

\section{Summary and discussion\label{Concluding Remarks}}
We have studied the competition of disorder and electron-electron interaction on honeycomb lattice. 
To this end, we have investigated the Anderson-Hubbard model with diagonal disorder at half-filling.
The analytic and local structure strong-coupling perturbation method which handles the Hubbard part of the Hamiltonian
allows us to address very large lattice sizes. 
To investigate the influence of interaction on disordered honeycomb lattice, we calculate the disorder-averaged DOS using KPM. 
Combined with the transfer matrix method in the ribbon geometry we are able to perform a careful finite size scaling
analysis which (being based on transfer matrix) is essentially free from finite size errors along the length of the ribbon.
For strong enough $U$ and $W$, our finite size scaling completely rules out the possibility of any metallic state 
between the Mott and Anderson insulating states. We therefore find a direct transition between Anderson and Mott states. 
The results indicate that the disorder shifts the Mott transition to larger values of the Hubbard $U$. In addition, the separation line of the Mott insulator and Anderson insulator is calculated from the criterion of vanishing the single particle gap which extracted from DOS. Two phases separate at $U \approx W$ for large interaction strengths.

Despite considering the lowest order perturbation theory -- which is already enough to get the Mott transition
in the clean limit -- our results agrees well with other numerical methods~\cite{Henseler,Atkinson,Srinivasan} in the
behavior or localization length in presence of the interaction. We found an interaction-induced enhancement of the localization length for weak and intermediate 
interaction strengths which is due to disorder screening. Although the localization length, is enhanced in this way, but still remains finite. 
For large interaction due to the suppression of hopping, the localization length decreases as the Mott localization starts to take over. It is curious to note
that at the onset of Mott insulation, the localization length becomes comparable to the lattice scale. This is similar to 
holographic description of the Mott phase which is identified as a phase where at the low-energy (IR) limit (i.e. near the Fermi level), the
system starts to recognize its UV lattice scale~\cite{Hartnoll}. 

Before ending the paper let us critically compare our finding of the absence of metallic phase between Mott and Anderson
insulating states presented in Fig.~\ref{phasediagram} with other published works which used the box distribution for disorder.
In Ref.~\onlinecite{Heidarian} by self-consistent Hartree-Fock calculations for a $50 \times 50$ lattice in two dimension a metallic phase which is 
sandwiched between Anderson insulator and Mott insulating state is obtained. In this reference
the physics of strong correlation (Mott transition) could not be addressed, as they used a mean field factorization of the Hubbard interaction
and hence they found a metallic state. 
In Ref.~\onlinecite{Srinivasan} the considerable influence of the Hubbard repulsion $U$ on delocalization was reported based on the results of the QMC method
for lattices consisting of up to $6\times 8$ sites. But owing to numerical restrictions, the Anderson insulator to metal transition in thermodynamic limit 
was not concluded as the maximum lattice size of $6\times 8$ was not conclusive to establish a metallic state in the thermodynamic limit.  
Possible metallic phase in between Anderson insulator and Mott insulator has been claimed by other studies based on QMC studies~\cite{Denteneer,Chakraborty} in two dimension as well as 
the results obtained from DMFT in infinite  dimension~\cite{Byczuk} at half-filling. Additionally, the dual-fermion approach in three dimension~\cite{Haase} 
showed the existence of the metallic phase in weak interaction strengths. All the above works have been done on lattices with small number of sites.
Thus the finite size effects on results are inevitable. The finite size effects become even more sever for honeycomb lattice where the localization 
length at small values of disorder is very large. 
On the other hand, a statistical DMFT study~\cite{Atkinson} on two-dimensional lattice with up to $1000$ sites did not find insulator-metal transition at strong disorder strengths which coincide with what we represented at Fig.~\ref{phasediagram}. Also, the authors of the Ref.~\onlinecite{Henseler} by the self-consistent study showed that existence of the metal phase is impossible.

\appendix

\section{One-point connected correlation function\label{cumulant}}

This Appendix gives the the one-point correlation function of the atomic limit of the Anderson-Hubbard model. We consider the unperturbed Hamiltonian $H_0$ of Eq.~(\ref{Hamiltonian}) as $H_0=\sum_{i} h_i$ where $h_i$ is expressed by,
\bea
h_i=U n_{i\uparrow} n_{i\downarrow}-\mu\ \sum_{\sigma}\ n_{i\sigma}+\sum_{\sigma}\ \epsilon_i\ n_{i\sigma}\label{atomiclimit}.
\eea
The one-point connected correlation function is defined as
\bea
\mathcal{G}_{i\sigma}(\tau,0)=-\langle\ T_\tau\ c_{i\sigma}(\tau)\ c_{i\sigma}^{\dagger}(0) \rangle,
\eea
where $T_\tau$ represent the time-ordering operator and the average is calculated with respect to local Hamiltonian $h_i$. Note that in the absence of magnetic field in the Hamiltonian ($\ref{atomiclimit}$), we can not distinguish between the one-point connected correlation function of the two spin projection $\sigma=\uparrow,\downarrow$. 
Let us rewrite the one-point connected correlation function as follow,
\bea
\mathcal{G}_{i\sigma}(\tau,0)=-\frac{1}{Z}\ \mathrm{Tr}\ \left(e^{-\beta h_i}\ c_{i\sigma}(\tau)\ c_{i\sigma}^{\dagger}\right),\nn
\eea
where the partition function $Z$ is given by,
\bea
Z=1+2e^{\beta(\mu-\epsilon_i)}+e^{\beta(2\mu-2\epsilon_i-U)}.
\eea
The one-point connected correlation function can be computed by inserting the identity operator,
\bea
\mathcal{G}_{i\sigma}(\tau,0)=-\frac{1}{Z}\sum_{nn'} \langle n|e^{-\beta h_i} e^{\tau h_i} c_{i\sigma} e^{-\tau h_i}|n'\rangle\langle n'|c_{i\sigma}^{\dagger}\ |n\rangle\nn
\eea
where the $|n\rangle$ and $|n'\rangle$ states denote the four possible states of a Hilbert space at each site, $|0\rangle, |\sigma\rangle, |\bar{\sigma}\rangle$ and 
$|\!\!\uparrow\downarrow\rangle$ which correspond to the empty, single occupied states with spin projection $\sigma$ and its opposite projection $\bar{\sigma}$ and double occupied state, respectively. The nonzero terms are given in the following,
\bea
\mathcal{G}_{i\sigma}(\tau,0)&=&-\frac{1}{Z}\ \langle 0|e^{-\beta h_i}\ e^{\tau h_i}\ c_{i\sigma}\ e^{-\tau h_i}|\sigma\rangle\langle \sigma|c_{i\sigma}^{\dagger}|0\rangle\nn\\
&-&\frac{1}{Z}\ \langle\bar{\sigma}|e^{-\beta h_i}\ e^{\tau h_i}\ c_{i\sigma}\ e^{-\tau h_i}|\uparrow\downarrow\rangle\langle \uparrow\downarrow|c_{i\sigma}^{\dagger}|\bar{\sigma}\rangle.\nn
\eea
So, we obtain,
\bea
\mathcal{G}_{i\sigma}(\tau,0)=-\frac{1}{Z}\Big( e^{\tau(\mu-\epsilon_i)}+e^{\beta(\mu-\epsilon_i)} e^{\tau(\mu-\epsilon_i-U)}\Big).
\eea
Fourier transformating to Matsubara frequencies we have,
\bea
\mathcal{G}_{i\sigma}(\mathrm{i}\omega)&=&\int_{0}^{\beta}\ d\tau\ e^{\mathrm{i}\omega\tau}\ \mathcal{G}_{i\sigma}(\tau,0)\nn\\
&=&\frac{1}{Z}\ \frac{1+e^{\beta(\mu-\epsilon_i)}}{\mathrm{i}\omega+\mu-\epsilon_i}\nn\\
&+&\frac{1}{Z}\ \frac{e^{\beta(2\mu-2\epsilon_i-U)}+e^{\beta(\mu-\epsilon_i)}}{\mathrm{i}\omega+\mu-\epsilon_i-U}.
\eea

By introducing the mean occupation $\langle n_{i}\rangle$ for each spin and lattice site $i$,
\bea
\langle n_{i}\rangle=\frac{e^{\beta(\mu-\epsilon_i)}+e^{\beta(2\mu-2\epsilon_i-U)}}{Z},
\eea
the one-point connected correlation function at arbitrary temperature $1/\beta$  becomes,
\bea
\mathcal{G}_{i\sigma}(\mathrm{i}\omega)=\frac{1-\langle n_{i}\rangle}{\mathrm{i}\omega+\mu-\epsilon_i}+\frac{\langle n_{i}\rangle}{\mathrm{i}\omega+\mu-\epsilon_i-U}.
\eea

At zero temperature limit or equivalently $\beta \to \infty$, the one-point connected correlation function is simplified to,
\bea
\mathcal{G}_{i\sigma}(\mathrm{i}\omega)&=&\frac{\Theta(\epsilon_i-\mu)}{\mathrm{i}\omega+\mu-\epsilon_i}+
\frac{1}{2}\frac{\Theta(\epsilon_i-\mu+U)\ \Theta(\mu-\epsilon_i)}{\mathrm{i}\omega+\mu-\epsilon_i}\nn\\
&+&\frac{\Theta(\mu-\epsilon_i-U)}{\mathrm{i}\omega+\mu-\epsilon_i-U}+
\frac{1}{2}\frac{\Theta(\epsilon_i-\mu+U)\ \Theta(\mu-\epsilon_i)}{\mathrm{i}\omega+\mu-\epsilon_i-U}.\nn\\
\eea

\section{Kernel polynomial method\label{KPM}}
Generally speaking, KPM is a numerical approach to calculate the spectral functions based on their expanding in Chebyshev polynomials~\cite{kpm,Habibi}. So, we can expand the DOS as follow,
\bea
    \hat{\rho}(\epsilon)=\frac{1}{\pi \sqrt{1-\epsilon^2}}\ \Big(\mu_0\  g_0+2\sum_{l=1}^{N_c}\mu_l\  g_l\  T_l(\epsilon) \Big),
    \label{dos-kpm} 
\eea
where $\epsilon$ is rescaled energy in such a way that fits in the range $[-1,1]$, $T_l(\epsilon)=\cos(l \arccos(\epsilon))$ is $l$'th Chebyshev polynomial,  
$g_l$s are the Jackson kernel coefficients which minimize the Gibbs oscillations and $\mu_l$ are Chebyshev moments. The sum is taken up to a cutoff number $N_c$.
It is important to note that in this method the Hamiltonian $H(E)$ with energy spectrum between $[E_{min},E_{max}]$ is rescaled to  $\hat{H}(\epsilon)$ where $ \hat{H}=(H-b)/a$, $\epsilon=(E-b)/a$, $b=(E_{max}+E_{min})/2$ and $a=(E_{max}-E_{min})/2$. Also, The moments are given by,
\bea
 \mu_l=\frac{1}{r} \sum_{r=1}^{M} \langle\phi_r\vert T_l(\hat{H}) \vert \phi_r \rangle,
 \eea
where $\phi_r$ are random single-particle states and $M$ is the number of random states used in numerical calculations. Furthermore, one can obtain the effect of $T_l(\hat{H})$ on a given ket using the recurrence relation of Chebyshev polynomials, namely, $T_l (\hat{H})= 2\hat{H} T_{l-1}(\hat{H}) - T_{l-2}(\hat{H})$ with initial conditions $T_{1}(\hat{H})=\hat{H}$ and $T_{0}(\hat{H})=1$.

To calculate the DOS for the Green's function in Eq.~\eqref{green'sfunction}, we use the following trick,
\begin{align}
    \rho'(\omega)=-\frac{1}{\pi}\lim_{\eta\to 0}\mathrm{Im}\frac{1}{E+i\eta+\Gamma^{-1}(i\omega)-V}\Big|_{E=0}.
\end{align}
Thus the Eq.~\eqref{dos-kpm} can be rewritten to,
\begin{align}
        \hat{\rho}'(\omega')=\frac{1}{\pi \sqrt{1-\epsilon^2}}\ \Big(\mu_0\  g_0+2\sum_{l=1}^{N_c}\mu_l(\omega')\  g_l\  T_l(\epsilon) \Big)\Big|_{\epsilon=0},
\end{align}
Where $\mu_l(\omega')$ are the generalized  KPM coefficients in which $H=\Gamma^{-1}(\omega)-V$. Also, $\omega'$ and $\hat{H}$ denote the rescaled    $\omega$ and $H$, respectively. To calculate $\mu_l(\omega')$, we need to compute $\mu_l$ for every $\omega'$ which is computationally expensive part of the calculations. 
So,  we used MPICH to parallel our program. Additionally, due to divergences of  $\Gamma^{-1}(\omega)$ for some values of disorder and making a large bandwidth, we set $N_c=15000$, $M=5$, and average it on $100$ configurations to obtain well converged values of DOS $\rho'(\omega')$ at $E=0$, .

\section{Transfer Matrix Method\label{TTM}}
In this appendix, we briefly explain the transfer matrix method used to calculate the localization length \cite{MacKinnon1,MacKinnon2}. 
The localization length $\lambda$ of the quasi-one dimensional system is defined as the characteristic length that specifies the 
exponential decay of wave function with the system length $L$~\cite{Janssen},
\bea
\psi(L)\ \propto\ \exp(-L/\lambda).
\eea
\begin{figure}[t]
    \centering
    \includegraphics[width=0.7\linewidth]{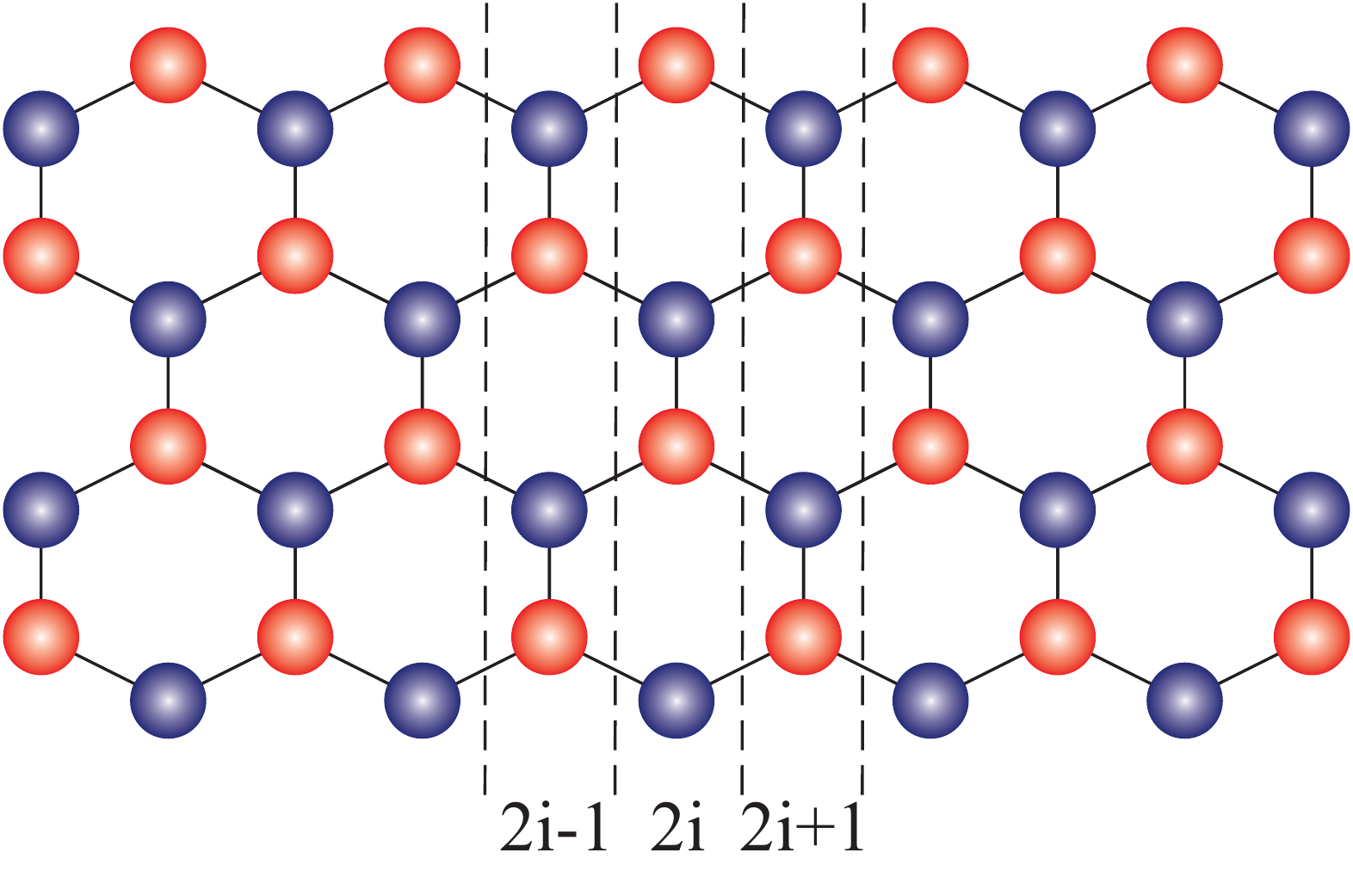}
    \caption{The honeycomb lattice used for transfer matrix method with transfer direction along zigzag edge and width $M=4$ and length $ L=11$.}
    \label{fig-latticeTM}
\end{figure}
In the transfer matrix method, the quasi-one dimensional Schr\"{o}dinger equation $\hat{H}\vec{\Psi }_i=E\vec{\Psi }_i$ is written as,
\bea
	\hat{V}_{i,i-1}^*\vec{\Psi }_{i-1}+\hat{H}_{i,i}\vec{\Psi }_i+\hat{V}_{i,i+1}\vec{\Psi }_{i+1}=E \vec{\Psi }_i .
\eea
So, the wavefunction $\vec{\Psi }_i$ of the $i$th slice along the transfer direction is calculated iteratively using the following transfer matrix equation,
\begin{align}
	\left(
	\begin{array}{c}
		\vec{\Psi }_{i+1} \\
		\vec{\Psi }_{i} 
	\end{array}
	\right) =
	\hat{T}_{i+1,i}
	\left(
	\begin{array}{c}
		\vec{\Psi }_{i} \\
		\vec{\Psi }_{i-1} 
	\end{array}\right),
\end{align}
where
\bea
\hat{T}_{i+1,i}=\left(
	\begin{array}{cc}
		\hat{V}^{-1}_{i,i+1}(E\mathbb{1}-\hat{H}_{i,i}) & -\hat{V}^{-1}_{i,i+1}\  \hat{V}_{i,i-1} ^*\\
		\\
		\mathbb{1} & \mathbb{0}\\
	\end{array}
	\right).
\eea
Here all the vector elements are $M \times M$ matrices and $T$ is $2M \times 2M$ matrix where $M$ denotes to the width of the system. Since the transport is stronger along the zigzag edge, in this work, we suppose the transport direction along this edge with periodic boundary condition as depicted in Fig.~\ref{fig-latticeTM}.

According to Oseledec’s theorem~\cite{Oseledec}, in the thermodynamic limit, the eigenvalues of
 \bea
\hat{\Gamma}=\lim_{N \to \infty}\left[\prod_{i=N}^1{\hat{T}^\dagger_{i+1,i}\prod_{i=1}^{N}{\hat{T}_{i+1,i}}}\right]^{1/{2N}},\label{Gamma_TM}
\eea
 converge to fixed values $e^{\pm\gamma_m}$ where $\gamma_m$ with $m=1,\cdots, M$ are Lyapunov exponents. The localization length is defined as the largest decaying length associated with the minimum Lyapunov exponent:
\bea
\lambda=\frac{1}{\gamma_\textrm{min}}.
\eea
Practically, to avoid numerical overflow, which came from multiplying the transfer matrices in Eq.~(\ref{Gamma_TM}), the Gram-Schmidt method is employed to orthonormalize the vectors. Let us note that we perform the  Gram-Schmidt orthonormalization after each multiplication due to severe fluctuations of the localization length on honeycomb lattice.
Additionally, in our calculation, $N$ is chosen in such a way that localization length converges.
 
\bibliographystyle{apsrev4-1}
\bibliography{Refs}
\end{document}